\documentclass[aps,prl,twocolumn,showpacs,amsmath,tightenlines]{revtex4}
\usepackage{graphics, bm}

\begin{document}

\title{Lifetime of metastable states in resonant tunneling structures}

\author{O. A. Tretiakov} \affiliation{Department of Physics, Duke
  University, Durham, NC 27708-0305}

\author{T. Gramespacher} \altaffiliation[Present address: ]{Winterthur
  Life, P.O. Box 300, CH-8401 Winterthur, Switzerland}

\author{K. A. Matveev} \affiliation{Department of Physics, Duke
  University, Durham, NC 27708-0305}

\date{September 19, 2002}  

\begin{abstract}
  We investigate the transport of electrons through a double-barrier
  resonant-tunneling structure in the regime where the current-voltage
  characteristics exhibit bistability.  In this regime one of the states
  is metastable, and the system eventually switches from it to the stable
  state. We show that the mean switching time $\tau$ grows exponentially
  as the voltage $V$ across the device is tuned from the boundary value
  $V_{th}$ into the bistable region. In samples of small area we find
  $\ln\tau\propto{|V-V_{th}|}^{3/2}$, while in larger samples
  $\ln\tau\propto{|V-V_{th}|}$.
\end{abstract}

\pacs{73.40.Gk, 73.21.Ac, 73.50.Td}

\maketitle

\begin{figure}
 \resizebox{.48\textwidth}{!}{\includegraphics{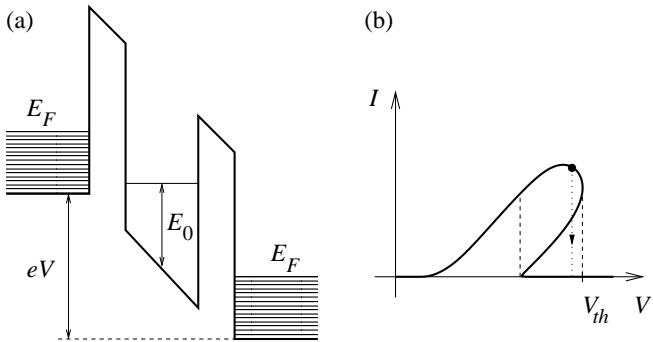}}
\caption{\label{fig1}(a) The potential profile of the DBRTS at applied 
  bias $V$. (b) The $I$-$V$ curve of the device has a bistable region
  between the dashed lines. The dotted line shows the process of switching
  from a metastable state to the stable one. }
\end{figure}

The problem of the decay of a metastable state has been addressed in a
variety of areas including first-order phase transitions~\cite{Langer},
Josephson junctions~\cite{Kurkijarvi}, field theory~\cite{Coleman},
magnetism~\cite{Victora}, chemical kinetics~\cite{Dykman1}. Meanwhile,
progress in nanofabrication technology has made possible observation of
intrinsic bistabilities in double-barrier resonant-tunneling structures
(DBRTS) \cite{Goldman:exp} and superlattices \cite{superlattice:1exp}.
Recent experiments \cite{Teitsworth, Grahn} with such devices have
demonstrated that near the boundary of the bistable region one of the two
states is metastable, and its lifetime has been studied by measuring
current as a function of time at different voltages. Thus, these devices
provide an ideal experimental system for studying the decay of metastable
states in real time. In this paper we develop the theory of switching
times in double barrier structures, Fig.~\ref{fig1}(a). We expect the
results to be relevant for other devices in which sequential resonant
tunneling plays a key role in describing the electronic transport, such as
weakly-coupled superlattices.

We concentrate on the case of intrinsic bistability, which can be observed
by measuring current $I$ as a function of voltage $V$ applied to the
device while the impedance of the external circuit equals zero. As shown
in Ref.~\onlinecite{Goldman:exp}, for a certain range of bias $V$, two
states of current $I$ are possible at the same value of the voltage, and
the $I$-$V$ curve has characteristic hysteretic behavior. As one increases
bias, the upper branch ends at some boundary voltage $V_{th}$, shown
schematically in Fig.~\ref{fig1}(b). If the voltage $V$ is fixed just
below the threshold $V_{th}$, the system stays in the upper state for a
finite time $\tau$, before decaying to the stable lower state.

\begin{figure}
 \resizebox{.35\textwidth}{!}{\includegraphics{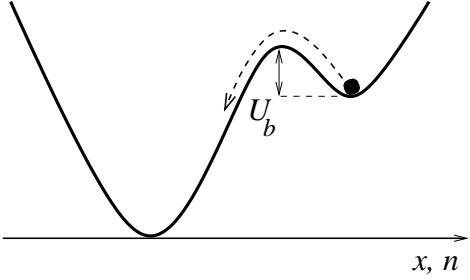}}
\caption{\label{fig2}Brownian particle in a double-well
  potential. The lifetime of the metastable state $\tau$ depends
  exponentially on the height of the barrier $U_b$ separating the local
  and global minima, $\tau\propto\exp(U_b/T^\ast)$. The coordinate of the
  particle $x$ has the meaning of the electron density $n$ in the well,
  and $U(n)=Su(n)$, see Eq.~(\ref{potential}).}
\end{figure}

We will show that the lifetime of the metastable state $\tau$ can be
understood by analogy to the problem of a Brownian particle in a
double-well potential (Fig.~\ref{fig2}). Here the coordinate of the
Brownian particle has the meaning of the current $I$ in the device (or the
electron density $n$). In the problem of the Brownian particle, $\tau$
depends exponentially on the height of the potential barrier $U_b$
separating the local and global minima, i.e.
$\tau\propto{\exp(U_b/T^\ast)}$, where $T^\ast$ is the temperature. Unlike
a Brownian particle, a DBRTS at nonzero bias is a non-equilibrium system
in which fluctuation phenomena are driven by shot noise in the current
rather than the electron temperature $T$. On the boundary of the bistable
region, the local minimum disappears, and therefore $U_b$ goes to zero.
Thus, it is clear that $\tau$ will depend exponentially on the voltage
measured from the boundary $V_{th}$ of the bistable region.

Here we investigate effects of shot noise in DBRTS using the framework of
the theoretical model introduced in Ref.~\onlinecite{Blanter99}. The DBRTS
is formed as a layered semiconductor heterostructure.  The electrostatic
potential across the device is shown in Fig.~\ref{fig1}(a). The potential
is assumed to be independent of the $x$ and $y$ coordinates. The model
includes only one subband in the quantum well. We furthermore assume that
at zero bias the bottom of this subband $E_0$ is above the Fermi energy
$E_F$ in the left and right leads. If the area of the sample $S$ is small,
we can assume that the charge in the well is distributed uniformly. Then,
the state of the device is completely described by the electron density
$n$ in the quantum well.  Below we will also discuss effects of
non-uniform charge distribution in the well, which are important in the
case of devices of large area.

In the sequential tunneling approximation, the transport in the device is
described by the following master equation for the time-dependent
distribution function $P(n,t)$ of the electron density $n$ in the well,
\begin{eqnarray}
\label{tunnelingMaster}
\frac{\partial P(n,t)}{\partial t}& =& P\left(n-\frac{1}{S},t \right)\sum_{{\bf q k}} W_{\bf k q}\left(n-\frac{1}{S}\right)f_{\bf k}(1-f_{\bf q})
\nonumber \\
&& +P\left(n+\frac{1}{S},t \right)\sum_{{\bf q p}} W_{\bf q p}\left(n+\frac{1}{S}\right)f_{\bf q}
\nonumber \\
&& -P(n,t)\sum_{{\bf q k}} W_{\bf k q}(n)f_{\bf k}(1-f_{\bf q})
\nonumber \\
&& -P(n,t)\sum_{{\bf q p}} W_{\bf q p}(n)f_{\bf q}.
\end{eqnarray}
Here $f_{\bf k}$, $f_{\bf p}$, and $f_{\bf q}$ are the Fermi occupation
numbers in the left lead, right lead, and the quantum well, respectively;
$W_{\bf k q}(n)$ and $W_{\bf q p}(n)$ are the tunneling rates through the
left and right barriers. The first two terms of
Eq.~(\ref{tunnelingMaster}) account for the processes which bring the
system to the state of electron density $n$, and the last two terms
describe the processes that take the system away from it. The first and
the third terms on the right-hand side of Eq.~(\ref{tunnelingMaster})
describe tunneling of one electron into the well from the left lead, while
the second and the fourth ones account for the probability of an electron
in the quantum well to tunnel into the right lead. We dropped the terms
describing the tunneling from the well to the left lead and tunneling from
the right lead into the well. These contributions are negligible because
the bistability emerges \cite{Blanter99} when the level in the well is
close to the bottom of the conduction band in the left lead and far above
the Fermi level in the right lead. (We assume $T\ll E_F$.)

Assuming that the total number of particles in the well is large, $nS\gg
1$, we can expand (\ref{tunnelingMaster}) in the small parameter $1/S$.
Keeping terms up to the second order we reduce the master equation to the
Fokker-Planck equation,
\begin{equation}
\label{tunnelingFPE}
\frac{\partial P(n,t)}{\partial t}=-\frac{\partial }{\partial n}[A(n)P(n,t)]+\frac{1}{2}\frac{\partial^2}{\partial n^2}[B(n)P(n,t)].
\end{equation}
The exact expressions for $A(n)$ and $B(n)$ are rather complicated, but
near the threshold they can be calculated analytically, see
Eq.~(\ref{AandB}) below.
 
The stationary solution of Eq.~(\ref{tunnelingFPE}) can be easily
obtained:
\begin{equation}
\label{potential}
P_{0}(n)=\frac{\mbox{const}}{B(n)}e^{-Su(n)},
\quad u(n)=-\frac{2}{S}\int_0^n \frac{A(n')}{B(n')}\,dn',
\end{equation}
where $u(n)$ is the effective potential.

In the derivation of the Fokker-Planck equation (\ref{tunnelingFPE}), the
coefficients $A(n)$ and $B(n)$ appeared as the first and second terms of
the expansion in $1/S$. We therefore conclude that $A(n)/B(n)\propto S$,
and $u(n)$ is independent of $S$. Thus when the area $S$ is large, the
distribution function $P_{0}(n)$ has very narrow peaks near the minima of
$u(n)$. If we neglect the fact that the width is finite, then the electron
density $n$ in the well can be found by minimizing $u(n)$. The
minimization condition written as $A(n)=0$ is in agreement with the
results of Ref.~\onlinecite{Blanter99}.

Our model allows for an analytical treatment at small values of the
parameter $\lambda={m{e}^2}/{2\pi{\hbar}^{2}C}$, where $C$ is the
capacitance of the device per unit area. Then the calculations are greatly
simplified, and one obtains the following expressions for $A$ and $B$ near
the threshold:
\begin{subequations}
\label{AandB}
\begin{eqnarray}
\label{A}
A(n)&=& -\frac{b}{2}\left[-\alpha+ \gamma (n-n_{th})^2\right],
\\
 \label{B}
b & \equiv & SB(n_{th})=\frac{\sqrt{2}}{\pi}\lambda^2 \frac{T_L^2}{T_R} \frac{C}{e^2}E_F^2,
\\
 \label{a0coefficient}
\alpha &=& 2\frac{1}{\lambda^2} \left(\frac{T_R}{T_L}\right)^2 \frac{E_0}{E_F^2}e\,\delta V,
\\
\label{b0coefficient}
\gamma &=& 2\frac{1}{\lambda^4} \left(\frac{T_R}{T_L}\right)^4 \left(\frac{e^2}{C}\right)^2 \frac{E_0^2}{E_F^4}.
\end{eqnarray}
\end{subequations}
Here $\delta V= V_{th}-V$, the electron density at the threshold
$n_{th}=\frac{\lambda^2}{2}\left(\frac{T_L}{T_R}\right)^2 \frac{C E_F^2}{e^2 E_0} $; and $T_{L,R}$ are the transmission
coefficients of the left and right barriers at energy $E_{0}$ and zero
applied bias. If $\lambda$ is not small, we cannot get explicit
expressions for $\alpha$, $\gamma$ and $b$, but the generic form of
Eq.~(\ref{A}) remains unchanged.

The potential $u(n)$ is shown schematically in Fig.~\ref{fig2} for a
voltage which lies slightly below the threshold voltage $V_{th}$. Close to
the threshold the potential can be approximated by a cubic polynomial,
\begin{equation}
\label{cubicPotential}
u(n)\approx -\alpha (n-n_{th})+\frac{\gamma}{3}(n-n_{th})^3+ u(n_{th}).
\end{equation}

Taking into account the exponential dependence of the mean switching time
on the barrier height, $\tau =\tau_{0}e^{U_b}$, and using
Eq.~(\ref{cubicPotential}), we find
\begin{equation}
\label{tauin1D}
\ln\frac{\tau}{\tau_{0}} =
\frac{4}{3}\frac{S\alpha^{3/2}}{\gamma^{1/2}}\propto S\, \delta V^{3/2}.
\end{equation}
The prefactor $\tau_{0}$ can be found using the techniques described,
e.g., in Ref.~\onlinecite{vanKampen}.

It is important to note that the form (\ref{cubicPotential}) of the
potential $u(n)$ and the linear dependence $\alpha\propto\delta V$ are
dictated by analyticity of the potential near the threshold. Thus, the
applicability of the following results is not limited to a particular
model of transport in DBRTS. A $3/2$-power law analogous to
Eq.~(\ref{tauin1D}) was theoretically predicted for different physical
systems in Refs.~\cite{Kurkijarvi, Victora, Dykman1, Dykman2}.
Experimentally it was observed recently for an optically trapped Brownian
particle~\cite{Dykman:exp}.

The result~(\ref{tauin1D}) has been obtained under the assumption that the
electrons spread rapidly in the $x$-$y$ plane, and their density $n$ is
uniform. In large samples, however, the spreading takes a long time, and
one has to account for the dependence of the density $n$ on the point
${\bf r}=(x,y)$ in the well. This can be done by generalizing the
Fokker-Planck equation~(\ref{tunnelingFPE}) to the case of non-uniform
${n({\bf r})}$.

We begin by studying the in-plane diffusion of electrons in the well
neglecting coupling to the leads. For simplicity we neglect the electron
correlation effects; the interactions of electrons will be accounted for
in the charging energy approximation. Assuming that the electrons diffuse
independently, one can write a master equation for the distribution
function $P\{n({\bf r}),t\}$ as follows. During the time $\Delta t$ at
most one electron can move from position ${\bf r}_{1}$ to ${\bf r}_2$,
that is,
\begin{eqnarray}
&&P\{n({\bf r}),t+\Delta t\}-P\{n({\bf r}),t\}= \int\!\!\int d{\bf r}_1  d{\bf r}_2
\nonumber \\
&&\times\Big[P\{n({\bf r})+\delta({\bf r}-{\bf r}_1)-\delta({\bf r}-{\bf r}_2),t\}W({\bf r}_{1},{\bf r}_2,\Delta t)  
\nonumber \\
&&-P\{n({\bf r}),t\} W({\bf r}_2,{\bf r}_{1},\Delta t) \Big].
\label{discreteP}
\end{eqnarray}
Here $W({\bf r}_{1},{\bf r}_2,\Delta t)$ is the probability density of an
electron diffusing from a point ${\bf r}_1$ in the plane of the quantum
well to point ${\bf r}_2$ during the time interval $\Delta t$. Since
electrons are fermions, a particle can diffuse only from a filled state at
${\bf r}_{1}$ to an empty state at ${\bf r}_2$. Assuming that the
diffusion is due to the elastic scattering of electrons by defects, we
find
\begin{equation}
\label{probabilityW}
W({\bf r}_{1},{\bf r}_2,\Delta t)= g({\bf r}_{1}-{\bf r}_2,\Delta
t)\nu \int f_{1}(E)[1-f_{2}(E)]\, dE, 
\end{equation} 
where $\nu$ is the density of states in the well (per unit area), and
$f_{1,2}(E)$ are the occupation numbers at points ${\bf r}_1$, ${\bf
  r}_2$. The classical diffusion probability $g(r,\Delta t)$ is given by
\begin{eqnarray}
\label{diffusion}
g({\bf r},\Delta t) = \frac{1}{4\pi D\Delta t}e^{-{\bf
r}^2/{4D\Delta t}}\simeq\delta({\bf r})+  D\Delta t\nabla^{2}\delta ({\bf r}),
\end{eqnarray} 
where $D$ is the diffusion coefficient. The approximate
form is obtained in the limit $\Delta t\to 0$.

Using Eqs.~(\ref{probabilityW}), (\ref{diffusion}) and expanding the
distribution function from Eq.~(\ref{discreteP}) up to the second order in
$\delta({\bf r}-{\bf r}_1)-\delta({\bf r}-{\bf r}_2)$ we obtain a
functional Fokker-Planck equation,
\begin{eqnarray}
\label{FokkerPlanck}
&& \frac{\partial P\{n({\bf r}),t\} }{\partial t} = \nu D\int\!\!\int d{\bf r}_1 d{\bf r}_2
\nonumber \\
&&\times \left[\left(\frac{\delta}{\delta n({\bf
      r}_1)}-\frac{\delta}{\delta n({\bf r}_2)}\right)+
\frac{1}{2}\left(\frac{\delta}{\delta n({\bf r}_1)}-\frac{\delta}{\delta
    n({\bf r}_2)}\right)^{2}\right]
\nonumber \\
&&\times\frac{\mu_{1}-\mu_{2}}{1-e^{-(\mu_{1}-\mu_{2})/T}} P\{n({\bf r}),t\}\nabla^{2}\delta ({\bf r}_{1}-{\bf r}_2).
\end{eqnarray} 
Here $\mu_{1}$ and $\mu_{2}$ are the electrochemical potentials at ${\bf
  r}_1$ and ${\bf r}_2$, respectively. Their values are found by adding
the electrostatic potential $e^{2}n/C$ to the Fermi energy $n/{\nu}$,
\begin{equation}
\label{mu1,2}
\mu_{1,2} =\frac{e^2}{\tilde C}n({\bf r}_{1,2}).
\end{equation} 
Here $\tilde C$ is defined by $e^{2}/{\tilde C}=e^{2}/{C}+1/{\nu}$.

Substituting Eq.~(\ref{mu1,2}) into Eq.~(\ref{FokkerPlanck}), integrating
twice by parts and assuming that $|\mu_1-\mu_2 |\ll T$, one obtains the
following Fokker-Planck equation,
\begin{equation}
\label{inplaneFPE}
\frac{\partial P\{ n,t\}}{\partial t}= -\nu
D\int d{\bf r} \frac{\delta}{\delta n} \left[\frac{e^2}{\tilde C}\nabla^2 n
+T\nabla^2\frac{\delta}{\delta n}\right] P\{ n,t\}. 
\end{equation} 
The stationary solution of Eq.~(\ref{inplaneFPE}) is found easily by
requiring the part of the integrand after the first functional derivative
to vanish,
\begin{eqnarray}
\label{P0inPlane}
P_{0}\{n\}= \exp\left[-\frac{1}{T}\int \frac{e^{2}n^2({\bf r})}{2\tilde
    C}\,d{\bf r}\right].
\nonumber
\end{eqnarray}
This result has a simple physical meaning. The equilibrium distribution
function $P_{0}$ has the Gibbs form $e^{-E/T}$, with the energy per unit
area ${e^{2}n^2}/{2{\tilde C}}$ in agreement with the electrochemical
potential (\ref{mu1,2}).

Using Eq.~(\ref{inplaneFPE}) we can take into account processes of charge
spreading in the quantum well. The electron density $n({\bf r})$ can
change due to either tunneling of electrons through the barriers or their
diffusion inside the well. Thus, we must add the terms from the right-hand
side of Eq.~(\ref{tunnelingFPE}) to Eq.~(\ref{inplaneFPE}) to account for
both processes. The combined Fokker-Planck equation takes the form
\begin{eqnarray}
\frac{\partial P\{ n({\bf r}) ,t\}}{\partial t}&=&
\int d{\bf r} \frac{\delta}{\delta n} \left[-A(n)
+\frac{S}{2} \frac{\delta}{\delta n}B(n) 
\right.
\nonumber \\
&&\left. -\nu D\frac{e^2}{\tilde C}\nabla^{2}n
\right] P\{ n({\bf r}),t\}.
\end{eqnarray}
Here we neglected the second term in Eq.~(\ref{inplaneFPE}). This can be
done as long as the temperature of the electrons in the well is much lower
than the Fermi energy~\footnote{More precisely, in the vicinity of the
  threshold \mbox{($\delta V\rightarrow 0$)}, one can show that this term
  is negligible at \mbox{$T\ll \lambda^{2}(1+2\lambda)({T_L}/{T_R})^3
    {E_F^3}/(E_0^{3/2}e\,\delta V)$}.}.

The stationary solution of this equation is
\begin{equation}
\label{P0general}
P_{0}\{n\}=\frac{\mbox{const}}{B(n)}e^{-F},\quad F\{n\}=\int d{\bf r}\left[u(n)+\eta (\nabla n)^{2} \right],
\end{equation}
where $\eta=\nu De^2/{\tilde C}SB(n)$. Note that in the limit $D\to\infty$
the electron density is uniform, $\nabla n=0$, and we recover the result
(\ref{potential}).

In the case when the diffusion coefficient $D$ is finite, the electron
density varies from point to point in the quantum well; hence, this
problem is infinite dimensional. In the multi-dimensional case, the system
escapes from the local minimum of potential through a point where the
barrier separating it from the global minimum takes the lowest possible
value, i.e., through a saddle point. The mean switching time $\tau$ is
determined by the potential at the saddle point, measured from the local
minimum. This approach is similar to the one used in the theory of
kinetics of first-order phase transitions \cite{Langer}, with $F$ playing
the role of the free energy.

The saddle point of the functional $F\{n\}$ is achieved at $n=n_s({\bf
  r})$ which almost everywhere in the sample is very close to the density
$n_{min}$ of the system at the local minimum of $F$.  However, in a region
of some characteristic size $r_0$, the density $n_s({\bf r})$ changes in
the direction of the global minimum, Fig.~\ref{fig2}. Thus, the DBRTS of
large area first switches to the stable state in a region of size $r_0$,
which then expands to the whole sample.

\begin{figure}[t]
\resizebox{.39\textwidth}{!}{\includegraphics{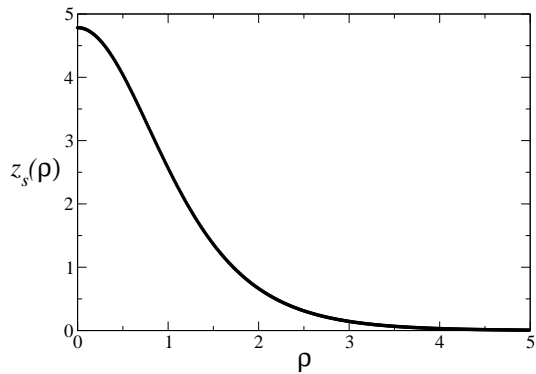}}
\caption{\label{fig3}Numerical solution of Eq.~(\ref{kink}). }
\end{figure}

We perform the following calculations in the regime of voltages very close
to the threshold $V_{th}$, where we can use $u(n)$ in the form
(\ref{cubicPotential}), and $\eta$ takes a constant value $\eta=\nu
De^2/{\tilde C}b$. Initially the system is in the local minimum, described
by a uniform density $n_{min}=n_{th}+\sqrt{{\alpha}/{\gamma}}$.  In order
to find the saddle point, it is convenient to parametrize the electron
density $n({\bf r})$ in terms of a dimensionless function $z(\bm{\rho})$,
such that $n({\bf r})=n_{min}-\sqrt{{\alpha}/{\gamma}}\, z({\bf
  r}/{r_0})$, where the length scale $r_0=\sqrt{2\eta}\,(\alpha
\gamma)^{-1/4}$. Using this parametrization, one can find the saddle point
of $F$ as a non-trivial solution $z_{s}(\bm{\rho})$ of the equation
\begin{equation}
\label{kink}
- \nabla^{2} z +2z-z^2=0,
\end{equation}
with the boundary condition $z_{s}(\bm{\rho})=0$ at $\rho\rightarrow
\infty$. It can be obtained numerically, Fig.~\ref{fig3}.

The switching rate $\tau^{-1}$ is proportional to the distribution
function $P_{0}\{n_{s}\}$, where $n_{s}({\bf
  r})=n_{min}-\sqrt{{\alpha}/{\gamma}}\, z_{s}({\bf r}/{r_0})$. Therefore,
one can calculate the logarithm of the mean switching time $\tau$ as
$F\{n_{s}\}-F\{n_{min}\}$,
\begin{equation}
\label{tau2}
\ln\frac{\tau}{\tau_{2}} = \kappa\frac{\alpha \eta}{\gamma}\propto \delta V.
\end{equation}
Here the constant $\kappa\simeq 62.01$ was found numerically~\footnote{
  The saddle point of a functional essentially identical to $F\{n\}$,
  Eqs.~(\ref{P0general}), (\ref{cubicPotential}), has been studied in
  Ref.~\onlinecite{Selivanov}. The numerical coefficient obtained in
  Ref.~\onlinecite{Selivanov} coincides with our result with 1\%
  accuracy.}.

According to Eq.~(\ref{tau2}), $\ln\frac{\tau}{\tau_{2}}$ does not
depend on the area $S$. On the other hand, since the critical fluctuation
$n_{s}({\bf r})$ can be centered anywhere in the sample, the switching
rate $\tau^{-1}$ is proportional to the area of the sample $S$, hence
$\tau_{2}\propto 1/S$. The exact calculation of the prefactor $\tau_{2}$
presents a number of theoretical challenges, which we leave for future
work.
  
In contrast to the case of small samples Eq.~(\ref{tauin1D}), the
logarithm of the escape time~(\ref{tau2}) in large samples is linear in
$\delta V$. The crossover between the two regimes occurs when the area $S$
of the sample is of the order of $r_0^2=2\eta(\alpha \gamma)^{-1/2}$. One
can see from Eq.~(\ref{a0coefficient}) that $r_0^2\propto \delta
V^{-1/2}$. Thus, one can observe this crossover in a single sample by
tuning the voltage. Indeed, at relatively small $\delta V$ we will have
$S\ll r_{0}^2$ and $\ln\tau\propto \delta V^{3/2}$, Eq.~(\ref{tauin1D}),
whereas at larger $\delta V$ we have $\ln\tau\propto \delta V$,
Eq.~(\ref{tau2}).
  
In conclusion, we have studied the switching time $\tau$ from the
metastable state to the stable one in DBRTS. We showed that $\tau$ is
exponential in the voltage measured from the boundary of the bistable
region; it is given by Eq.~(\ref{tauin1D}) or (\ref{tau2}) depending on
the area of the sample. Our results can be tested in experiments similar
to Refs.~\onlinecite{Teitsworth, Grahn}.

We are grateful to M.~I.~Dykman, H.~T.~Grahn and S.~W.~Teitsworth for
valuable discussions. T. G. acknowledges support from the Swiss National
Science Foundation. This work was also supported by the NSF Grants
DMR-9974435 and DMR-0214149 and by the Sloan foundation.

\end{document}